\documentclass[12pt,a4paper]{amsart}
\newtheorem{de}{Definition}[section]

\newtheorem{re}[de]{Remark}

\newtheorem{te}[de]{Theorem}

\usepackage{soul}
\usepackage{amsfonts} 
\usepackage{amsmath}
\usepackage[all]{xy}

\newcommand{\ot}{\otg{B}}

\newcommand{\mproof}{\noindent{\bf Proof.}}
\newcommand{\twosid}[3]{\ar@<0.25ex>@{<-}[#1]^{#2}  \ar@<-1ex>[#1]_{#3}}

\newenvironment{Proof}{{\mproof}}{{ }\hfill$\Box$ \newline\vspace{\lineskip}}

\def\ot{\otimes}

\begin{document}

\title{Yang-Baxter Equations, Computational Methods\\
 and Applications}
\maketitle
\begin{center}
\author{Florin F. Nichita}\\
{\it Simion Stoilow Institute of Mathematics
of the Romanian Academy\\
21 Calea Grivitei Street, 010702 Bucharest, Romania}
\end{center}

%\markboth{Journal of \LaTeX\ Class Files,~Vol.~6, No.~1, January~2007}%
%{Shell \MakeLowercase{\textit{et al.}}: Bare Demo of IEEEtran.cls for Journals}
%
\bigskip

\bigskip

%--------------------------------------------------------------------------------------------------------------
{\bf Abstract}\\ 
Computational methods are an important tool for solving the Yang-Baxter equations
(in small dimensions), for classifying (unifying) structures, and for solving related problems.
This paper is an account of some of the latest developments on the Yang-Baxter equation,
its set-theoretical version, 
and its applications. 
We construct new set-theoretical solutions for the Yang-Baxter equation.
Unification theories and other results are  proposed or proved.
%\end{abstract}

%\bigskip

\bigskip
%\begin{IEEEkeywords}


%------------------------------------------------------------------------------------------------------------

%\IEEEpeerreviewmaketitle

\section{Introduction}

The current paper is an extension 
of \cite{rfm}, a paper  based on a presentation at the INASE conference in Barcelona. 
Our interaction with the participants and some earlier proceedings of INASE
influenced the
development of it.

Computational methods were an important tool for solving
the Yang Baxter equation and Yang-Baxter system, in dimension two, in the papers \cite{hietarinta, HlaSno:sol}, where
the authors used Grobner basis.

The discovery of the Yang-Baxter equation (\cite{py}) in theoretical physics and statistical mechanics
(see \cite{yang, baxter1, baxter2}) has led to many applications in these fields and in 
quantum groups, quantum computing, knot theory, braided categories, analysis of integrable systems, quantum mechanics, etc (see \cite{si}).
The interest in this equation is growing, as new properties of it are found, and
its solutions are not classified yet (see also \cite{dpfn, hin}).

One of the striking properties of this equation is its
unifying feature (see, for example, \cite{fnichita, lebed, lebed2}).
Another unification of non-associative structures, was recently
obtained using the so-called UJLA structures
(\cite{rfm, inn, inn2}), which 
could be seen as structures which comprise the information
encapsulated in associative algebras, Lie algebras and Jordan algebras.
These unifications have similarities with the properties
of quantum computers, whose quantum gates
can execute many operations
in the same time (a classical gate executes just
one operation at a time). 
Several Jordan structures have applications in quantum group theory and
exceptional Jordan algebras play an important role in recent
fundamental physical theories
 namely, in the theory of super-strings 
(see \cite{I}).

The quantum computer can be used to solve large computational
problems 
from number theory and optimization.
An example is the Shor's algorithm, a quantum algorithm that determines
quickly and effectively the prime factors of a big number.
With enough qubii, such a computer could use the Shor's algorithm
to break algorithms encryption used today.

%The importance of computers in our days is that big that
%we could call our times the ``computers era''.
%The quantum mechanics had an important influence
%on building computers; for example, it led to the production of  transistors.
%At present, the quantum mechanics laws are used for the processing and transmission of information.
%The first quantum computer (which uses principles of quantum mechanics) was sold to the aerospace and security of defense company Lockheed Martin.
%The first quantum computer (which uses principles of quantum mechanics) was sold to the aerospace and security of defense company Lockheed Martin.
%for 10 million dollars.  
%The manufacturing company, D-Wave, founded in 1999 and called ``a company of quantum computing" promised to perform professional services for
%the computer maintenance as well.

%\bigskip

%\bigskip

The organization of our paper is the following. 
In the next section we give some preliminaries on the Yang-Baxter equation, and we explain
its importance for constructing quantum gates and obtaining link invariants.
%\bigskip
Section 3 deals with 
the set-theoretical Yang-Baxter equation.
New solutions for it are presented.
Section 4 deals with transcendental numbers, computational methods and some  applications.
In Section 5, we discuss about
algorithms and interpretations 
of the Yang-Baxter equation in computer science.
%\bigskip
Section 6  is about unification theories for non-associative algebras, and their connections
with the previous sections. 
A conclusions section ends our paper.

\section{Yang-Baxter equations}
%, knots and quantum gates}

The Yang-Baxter equation first appeared in theoretical physics, in a paper by the Nobel laureate C.N. Yang, and in statistical mechanics, in R.J. Baxter's work. 
It has applications in many areas of physics, informatics and mathematics.
Many scientists have used computer calculations or
the axioms of various algebraic structures
in order to solve this equation, but
the full classification of its solutions remains an open problem (see \cite{rfm, dpfn, si, vd, a7, a9, DasNic:yan, nipo, nonlin}).

\bigskip

In this paper tensor products are defined over the field $k$.

For $ V $ a $ k$-space, we denote by
$ \   \tau : V \otimes V \rightarrow V \ot V \  $ the twist map defined by $ \tau (v \ot w) = w \ot v $, and by $ I: V \rightarrow V $
the identity map of the space $V$;
for $ \  R: V \ot V \rightarrow V \ot V  $
a $ k$-linear map, let
$ {R^{12}}= R \ot I , \  {R^{23}}= I \ot R , \
{R^{13}}=(I\ot\tau )(R\ot I)(I\ot \tau ) $.

%\bigskip

\begin{de} A
{ \it  Yang-Baxter
operator} is $ k$-linear map
$ R : V \ot V \rightarrow V \ot V $,
which is invertible, and it satisfies the braid condition (sometimes called the {\em Yang-Baxter equation}):
%is defined as an invertible  $ k$-linear map  $ R : V \ot V \rightarrow V \ot V $
%which satisfies the  equation:
\begin{equation}  \label{ybeq}
R^{12}  \circ  R^{23}  \circ  R^{12} = R^{23}  \circ  R^{12}  \circ  R^{23}.
\end{equation}
If $R$ satisfies (\ref{ybeq}) then both
$R\circ \tau  $ and $ \tau \circ R $ satisfy the {\em quantum Yang-Baxter equation} (QYBE):
\begin{equation}   \label{ybeq23}
R^{12}  \circ  R^{13}  \circ  R^{23} = R^{23}  \circ  R^{13}  \circ  R^{12}.
\end{equation}
\end{de}
Therefore, the equations (\ref{ybeq}) and (\ref{ybeq23}) are equivalent.

%\bigskip
%\begin{re} \label{alg}
For $A$ be a (unitary) associative $k$-algebra, and $ \alpha, \beta, \gamma \in k$, the authors of
  \cite{DasNic:yan} defined the
$k$-linear map
$ R^{A}_{\alpha, \beta, \gamma}: A \ot A \rightarrow A \ot A, $
\begin{equation} \label{ybdn}
 a \ot b \mapsto \alpha ab \ot 1 + \beta 1 \ot ab -
\gamma a \ot b 
\end{equation}
which is a Yang-Baxter operator if and only if one
of the following cases holds:\\ 
(i) $ \alpha = \gamma \ne 0, \ \ \beta \ne 0 $; $ \ $
(ii) $ \beta = \gamma \ne 0, \ \ \alpha \ne 0 $; $ \ $
(iii) $ \alpha = \beta = 0, \ \ \gamma \ne 0 $.
%\end{re}

%\begin{re}
  
% \end{re}
An interesting property of (\ref{ybdn}), can be visualized in knot theory, where
the  link invariant associated to $ R^{A}_{\alpha, \beta, \gamma}$
is the Alexander polynomial (cf. \cite{T, mn}).

For $ ( L , [,] )$  a Lie super-algebra over $k$,
  $ z \in Z(L) = \{ z \in L : [z,x]=0 \ \ \forall \ x \in L \} ,
 \ \vert z \vert =0 $ and $ \alpha \in k $, the authors of the papers \cite{mj} and \cite{nipo}  defined the following Yang-Baxter operator:
${ \phi }^L_{ \alpha} \ : \ L \ot L \ \ \longrightarrow \ \  L \ot L $,
\begin{equation} \label{Lie}
x \ot y \mapsto \alpha [x,y] \ot z + (-1)^{ \vert x \vert \vert y \vert } y \ot x \ .
\end{equation}

This construction could lead to interesting ''bozonisation'' constructions,
a technique often used in constructing (super) quantum groups.
For example, the FRT algebras associated to ${ \phi }^L_{ \alpha}$
for a Lie algebra and a Lie super-algebra might be related
via such a construction.

\begin{re} 
In dimension two, $ R^{A}_{\alpha, \beta, \alpha} $ gives
a universal quantum gate (see \cite{rfm}):
\begin{equation} \label{rmatcon5}
\begin{pmatrix}
1 & 0 & 0 & 0\\
0 & 0 & 1 & 0\\
0 & 1  & 0 & 0\\
0 & 0 & 0 & -1
\end{pmatrix}
\end{equation}
which, according to \cite{kl}, 
is related to the CNOT gate:

\begin{equation} \label{rmatcon3}
CNOT=
\begin{pmatrix}
1 & 0 & 0 & 0\\
0 & 1 & 0 & 0\\
0 & 0  & 0 & 1\\
0 & 0 & 1 & 0
\end{pmatrix}
\end{equation}
\end{re}

\begin{re}
The author of the paper \cite{zeng} obtains
the abstract controlled-not by a composition
of a comonoid and a monoid.
That formula could be related to the well-known
Yang-Baxter operator
$ \ \  a \ot b \mapsto \sum a_1 \ot a_2 b $,
and leads us to the open problem of comparing this
operator with (\ref{ybdn}).

\end{re}

\begin{re}
The matrix (\ref{rmatcon5}) can be interpreted as a sum of the Yang-Baxter operators $I$ and $-I$, using
the techniques of \cite{bn}.

\end{re}

%\bigskip

\section{The set-theoretical Yang-Baxter Equation}

\begin{de} 
For an arbitrary set $X$, the map
$ S : X \times X \rightarrow X \times X $,
is a solution for
the
{ \it set-theoretical Yang-Baxter equation
} if
\begin{equation}   \label{ybeq234}
S^{12}  \circ  S^{13}  \circ  S^{23} = S^{23}  \circ  S^{13}  \circ  S^{12}.
\end{equation}
(Here $S^{12}= S \times I, \  S^{23}= I \times S $, etc.)

\end{de}

There are many examples of solutions for the equation (\ref{ybeq234}): from ``brace'' structures,
from relations on sets, etc,
and they are related to other interesting structures
(see, for example, \cite{stybe, tgi, a9}).
However, this equation is not completely solved yet.

\bigskip

Next, we present some explicit solutions for (\ref{ybeq234}),
we extend
some constructions from \cite{rfm}, and then we give  new constructions
of solutions for (\ref{ybeq234}).

\bigskip

We
consider 
a three dimensional Euclidean space, and a point $ P(a,b,c)$ of it.

The symmetry of the point $P(a,b,c)$ about the origin is defined as follows:

$S_{O} (a,b,c) = (-a, \ -b, \ -c) $.

The symmetries of the point $P(a,b,c)$ about the axes $OX, \ OY, \ OZ$ are defined as follows:

$S_{OX} (a,b,c) = (a, \ -b, \ -c)$,

$  S_{OY} (a,b,c) = (\ -a, \ b, \ -c)$,

$  S_{OZ} (a,b,c) = (\ -a, \ -b, \ c)$.

%\bigskip

The symmetries of the point $P(a,b,c)$ about the planes $XOY, \ XOZ, \ YOZ$ are defined as follows:

$S_{XOY} (a,b,c) = ( a, \ b, \ - c)$,

$S_{XOZ} (a,b,c) = ( a, \ -b, \  c)$,

$ S_{YOZ} (a,b,c) = (\ -a, \ b, \  c)$.

%\bigskip 

The above symmetries with the identity map form a group:
 
$ \{ I, S_{OX},  S_{OY},  S_{OZ},  S_{XOY},  S_{XOZ},  S_{YOZ},  S_O \}$,

which contains a subgroup isomorphic with Klein's group:
$ \{ I, S_{OX},  S_{OY},  S_{OZ}  \}$.

One could check the following instances of the Yang-Baxter equation:

%\begin{equation} \label{sim}
$S_{XOY} \circ S_{XOZ} \circ S_{YOZ} = S_{YOZ} \circ  S_{XOZ} \circ S_{XOY}$,
%\end{equation}
%\begin{equation} \label{sim2}

$S_{OX} \circ S_{OY} \circ S_{OZ} = S_{OZ} \circ  S_{OY} \circ S_{OX}$.
%$\end{equation} 

\begin{te} \label{t1}
The following is a two-parameter family of solutions for the
set-theoretical Yang-Baxter equation:

$S: \mathbb{R} \times \mathbb{R} \rightarrow
 \mathbb{R} \times \mathbb{R} $,
$ \ \ \ (x, y) \mapsto  (y^{\alpha}, \ x^{\beta} y^{1 - \alpha \beta}) \ \ \ 
\forall \alpha, \beta \in \mathbb{N^*} $.
\end{te}

\begin{Proof}
 We observe that the map
$ (x, y) \mapsto (x^m y^n, x^p y^q)$
 is a solution for (\ref{ybeq234})
if and only if the following relations hold:

$ mnq=0, \ mpq=0, \ m q^2 = m^2 q,$
$ m^2 + mnp= m, \ q^2 + npq= q $.

We leave the analysis of this system of non-linear equations as a computational problem, and we pick just the above solution of this system. 

Another approach to study this theorem might be by using
the properties of generalized means from \cite{ng}. 
\end{Proof}

\begin{te} \label{t2}
The following is a two-parameter family of solutions for the
set-theoretical Yang-Baxter equation:

$R: \mathbb{C} \times \mathbb{C} \rightarrow
 \mathbb{C} \times \mathbb{C} $,
 $ \ \ \ (z, w) \mapsto  ({\alpha} w , \ {\beta} z + ({1 - \alpha \beta}) w) \ \ \ 
\forall \alpha, \beta \in \mathbb{C} $.
\end{te}
\begin{Proof} One way to prove this theorem
is to follow the steps of the above proof.
\end{Proof}

\begin{re} Another way to prove Theorem \ref{t2}
is to relate it to Theorem \ref{t1}.
Thus, in some cases,
 the exponetial function,
$ f: \mathbb{C}  \rightarrow
 \mathbb{C}, \ \ z \mapsto e^z, $
is a morphism of the above solutions for (\ref{ybeq234}):
$\ \ \ \ (f\times f)\circ R \ = S \circ
(f\times f) \  .$  

Further, approaches could be by using
computational methods, and extended results
for Dieudonne modules 
(see \cite{sa}) are expected.
 
\end{re}

\begin{te}
The following is a  solution for the
set-theoretical Yang-Baxter equation:

$S': \mathbb{R^*} \times \mathbb{R^*} \rightarrow
 \mathbb{R^*} \times \mathbb{R^*} $,
$ \ \ \ \ (x, y) \mapsto (\frac{x}{y}, x^{2}) $.
\end{te}

\begin{Proof} The direct proof is the shortest.
Notice the relationship of $S'$ with $S$
from Theorem \ref{t1}.
\end{Proof}

Other examples of solutions for (\ref{ybeq234})
will be given in the Section 5, and they will
be related to informatics.

\section{Transcendental numbers and applications}

The following identity, containing the transcendental
numbers $ \ e $  and $ \ \pi $ is well-known (see more details about transcendental numbers in \cite{solnic}):

\begin{equation} \label{eipi}
e^{i \pi} + 1= 0  .
\end{equation}

Let $J$ be  the following matrix  (for $ \alpha \in \mathbb{R}^*$):
\begin{equation} \label{rmatcon2}
\begin{pmatrix}
0 & 0 & 0 & \frac{1}{\alpha} i\\
0 & 0 & i & 0\\
0 & i & 0  & 0\\
\alpha i & 0 & 0 & 0
\end{pmatrix}
\end{equation}
then, similarly to (\ref{eipi}), the following formula holds: 
$$ \ \ e^{\pi \ J} + I_4 = 0_4 \ \ \  \ \ \ \  J, \ I_4 , 0_4 \in \mathcal{M}_4 ( \mathbb{C}) \ . $$ 

$R (x) = \cos x I_2 + \sin x J= e^{xJ}: V^{\ot 2} \rightarrow 
 V^{\ot 2} $
satisfies the colored Yang-Baxter equation:
\begin{equation} \label{ybco}
R^{12}(x) \circ  R^{23}(x+y)  \circ  R^{12}(y) \ 
 = \ R^{23}(y)  \circ  R^{12}(x+y)  \circ  R^{23}(x) \ \ . \
\end{equation}

Also, $R(x)= e^{xJ}$ is a solution for the following differential
matrix equation:

\begin{equation} \label{diff}
 Y' = J Y \ ,
\end{equation}

which is related, for example, to \cite{mef}. The combination
of the properties (\ref{ybco}) and (\ref{diff}), has applications
in computing the Hamiltonian of many body systems in
physics.

Computational methods could be employed for
finding matrices $J$ with these properties in higher dimensions. 

\bigskip

The presentation \cite{ff} was related to the
following formula with transcendental numbers:
$\int^{+ \infty}_{- \infty} e^{- x^2} dx = \sqrt{\pi} \ .$ 
Thus, the experimental results presented at that time were related
to the Gauss bell function.

\bigskip

Next we solve an open problem
proposed in \cite{rfm}. This theorem, which is related to
the transcendental numbers $e$ and $\pi$, was solved, using
computational methods,
thanks to an observation of Dr. Mihai Cipu.

\begin{te} \label{N}
$ \sum^n_1 \frac{1}{k^2} < \frac{2}{3}  \ ( \frac{n+1}{n} )^n \ \ \ \ \forall n \in \mathbb{N^*} . $ 
\end{te}

\begin{Proof}
We evaluate the first three values for the above expressions:
\begin{center}
  \begin{tabular}{ | l || c || r| }
    \hline
    n=1 & 1 & 1.(3) \\ \hline
    n=2 & 1.25 & 1.5 \\ \hline 
    n=3 & 1.36(1) & 1.58... \\ 
    \hline
  \end{tabular}
\end{center}
The inequality is true in these cases,
and we will  use the following inequality:
$ \sum^{n}_1 \frac{1}{k^2} < \frac{\pi^2}{6} < 1.645     \ \  \ \forall n \geq 4$,
in order to finish our proof.
The above inequality is a consequence of the Basel problem:
$ \sum^{\infty}_1 \frac{1}{k^2} = \frac{\pi^2}{6} $.
(A proof for this identity was recently given in \cite{basel}.)
Note also that the sequence in the right hand side is
increasing.
The last step of the proof is shown bellow.
\begin{center}
  \begin{tabular}{ | l || c || r| }
    \hline
   n= 4 &  1.4236(1) & 1.6276... \\ \hline
    n= 5 & ...  &  1.65(8) \\ \hline \hline \hline
   $ \infty$ & 1.644934067... & 2.718281828... \\
    \hline
  \end{tabular}
\end{center}
\end{Proof}

%\bigskip

Other recent problems relating  $ \ e $  and $ \ \pi $  are listed below.
Numerical and experimental results are very important in studying them.

%\begin{equation} \label{new6}
$$ \mid e^{1-z} + e^{ \bar{z}} \mid > \pi \ \ \  \ \ \ \forall z \in \mathbb{C} \ , \ $$
%\end{equation}
$$  \int^{b}_{a} e^{- x^2} dx \  < \  \frac{e^e}{\pi} ( \frac{1}{e^{\pi a}}  -  \frac{1}{e^{\pi b}}) \ , \ \ \forall  \ a,b \in \mathbb{R}, \ a \leq b, $$

$$  x^2 + e >  \pi x  \ \ \ \forall x \in \mathbb{R}; $$
the last inequality holds because
$ \Delta= \pi^2-4e = - 1, 003522913... < 0 \ .$
We conjecture that $ 4e-  \pi^2 = 1,003522913... $ is a transcendental number.
Theorem \ref{N}, and
numerical results could give a partial answer for this problem.

%\bigskip

The geometrical interpretation of the formula 
$\ \pi^2  < \ 4  e  \ $
could be stated as:
``The length of the circle with diameter $\pi$ is almost equal (and less) to the perimeter of a square with edges of length $e$''.

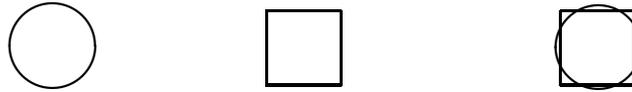
\begin{picture}(220,70)
%\begin{center}
\thicklines
\put(100,40){\circle{31.4}}
\put(180,25){\line(0,1){28}}
\put(180,25){\line(1,0){28}}
\put(180,53){\line(1,0){28}}
\put(208,25){\line(0,1){28}}
\put(304.3,39.3){\circle{31.4}}
\put(290,25){\line(0,1){28}}
\put(290,25){\line(1,0){28}}
\put(290,53){\line(1,0){28}}
\put(318,25){\line(0,1){28}}
\end{picture}
%}

The area of the above circle is greater than the area of the square, because
$ \pi^3 > 4 e^2 $.

%\begin{figure}[h!]
 % \caption{{A circle with diameter $\pi$  and a square with edges of length $e$
%have almost the same perimeter.}}
 %{ \centering

%\end{figure}

\bigskip

{ \it
OPEN PROBLEMS. For an arbitrary convex closed curve, we consider the largest diameter (D). (It can be found by considering the center of mass of a body which
corresponds to the domain inside the
given curve.)

(i) The equation 
\begin{equation} \label{ecgraddoi}
x^2 - \frac{L}{2} x + A =0 
\end{equation}
and its implications are not completely understood. For example,
if the given curve is an ellipse, solving this equation in terms of the semi-axes of the ellipse is an unsolved problem. 

(ii) We conjecture that the following system of equations  is an  
inverse of (\ref{ecgraddoi}). We consider two
functions with second order derivatives, such that}

$ f : [0, D]  \rightarrow \mathbb{R}, \ f \geq 0, \ f'' \leq 0, \ \
 g : [0, D]  \rightarrow \mathbb{R},  \ g \leq 0, \ g'' \geq 0, $\\
$ \int^D_0 \sqrt{1+ (f'(x))^2} + \sqrt{ 1 + (g(x)')^2} \  dx \ = L \ ,  \   \ 
 \int^D_0 f(x)  -g(x)  \ dx \ = A \ . $ 
%\end{equation}
%}

%\bigskip

\begin{re}
 
Graphics for arbitrary closed convex curves related to  the above open-problems could be
represented using graphing calculators and computers.  Thus, some numerical (experimental) results
can be obtained.
This direction seems to be a challenging one for computer scientists and it has applications
for  representations similar to those from \cite{maf}.  
\end{re}

\bigskip

\begin{re}
The equation 
$ x^i = i^x  $ for  $  \ \ x \in \mathbb{R^*_+} \  $,
has no solutions (see \cite{rfm}).

 At this moment we do not have convincing
 numerical / experimental results for solving the following generalization of
the above equation:
$ \ \  z^i = i^z \ \ \ \ \ z \in \mathbb{C^*} \ . $

\end{re}

\bigskip

\section{The Yang-Baxter Equations in Informatics}

The Yang-Baxter equation  
represents some kind of compatibility condition
in logic.

More explicitly,
let us consider  three logical sentences $p, \ q, \ r$, and let us suppose that
if all of them are true, then the conclusion A could be drawn, and if $p, \ q, \ r$ are all false then the conclusion C can be drawn.

Modeling this situation by the map $R$, defined by $ ( p, \ q) \ \mapsto (p'= p \vee q, \ q'= p \wedge q) $,
helps us to comprise our analysis: we can apply $R$ to the pair $(p, \ q)$, then to $(q', \ r)$, and, finally to
$(p', \ q'')$.

The Yang-Baxter equation guarantees that the order in which
we start this analysis is not important; more explicitly, in this case,
it states that $ \ ( \ (p')', \ q''', \ r') \ = \ ( \ p', \ q''', \ (r')') \ $.
In other words, the map $ ( p, \ q) \ \mapsto ( p \vee q, \ p \wedge q) $
is a solution for (\ref{ybeq234}).
%then, it is not important the order in which we apply this operator to  $p, \ q, \ r$. 

\bigskip

%Another interpretation of the Yang-Baxter equation is related to the algorithms which order sequences of numbers.
The sorting of numbers (see, for example, \cite{sorting, rfm}) is an important problem in informatics, and the Yang-Baxter equation is related
to it.
The following ``Bubble sort'' algorithm is related to the right hand side of (\ref{ybeq}).

%\bigskip

%\mathscr{
{\bf int} m, aux;\\
m=L;\\
{\bf while} (m)\\
$\{$\\
{\bf for} (int i=1; i$\le$L-1; i++)\\
{\bf if} (s[L-i] $\ge$ s[L+1-i])\\
$\{$\\
aux = s[L+1-i];\\
s[L+1-i] = s[L-i];\\
s[L-i] = aux;\\
$\}$\\
m - -;\\
$\}$\\
%}\\
The mai part of another sorting algorithm, related to the left hand side of (\ref{ybeq23}), is given below (see \cite{rfm}).

\bigskip

           {\bf if} (s[i]$\ge$s[j])\\
           $\{$\\
                        aux=s[i];\\
                        s[i]=s[j];\\
                        s[j]=aux;\\
           $\}$\\
%-------------------------------------------------\\
%$ \cdots$

%\newpage

Ordering three numbers is related to a common solution of (\ref{ybeq}) and (\ref{ybeq23}):
$ R (a, b) = (min(a,b), max(a,b)) $.
In a similar manner, one can finding the greatest common divisor and the least common multiple
of three numbers,
using another common solution of (\ref{ybeq}) and (\ref{ybeq23}):
$ \ R' (a, b) = (gcd(a,b), lcm(a,b)) $.

Since $R$ and $R'$ can be extended to braidings in  certain monoidal categories, we obtain interpretations for the cases when we 
deal with more numbers. 
The ``divide et impera'' algorithm for finding the maximum / minimum (or the
greatest common divisor /
least common multiple) of a sequence of numbers could be related to Yang-Baxter systems and to the gluing procedure from \cite{bn}.

\bigskip

{\centering\section{Non-associative Algebras and their unifications}}

The main non-associative structures are Lie algebras and Jordan algebras.
Arguable less known, Jordan algebras have applications
in physics, differential geometry, ring geometries, quantum groups,
analysis, biology, etc (see \cite{I, RI, RIordanescu}).

 The formulas (\ref{ybdn}) and (\ref{Lie}) lead to the unification of
associative algebras and Lie (super)algebras in the framework 
of Yang-Baxter structures (see \cite{humbold,nichita}). 
For the invertible elements of a Jordan algebra,
one can associate a symmetric space, and, after that,
a Yang-Baxter operator.
Thus, the Yang-Baxter equation can be thought as a unifying equation.

Another unification for these structures will be presented below.

\bigskip

\begin{de}
For the vector space V, let 
   $ \eta : V \otimes V \rightarrow V, \ \ 
\eta (a \otimes b) = ab ,  $  be a linear map which satisfies:
\begin{equation} \label{new}
 (ab)c + (bc)a + (ca)b  = a(bc) + b(ca) + c(ab),
\end{equation} 
\begin{equation} \label{Jordan}
 (a^2 b) a \ = \ a^2 (ba), \ 
 (a b) a^2 \ = \ a (b a^2), \
 (b a^2) a \ = \  (ba) a^2 , \
 a^2 (ab) \ = \  a (a^2 b),
\end{equation} 
$\forall \ a, b, c \in V $.

Then, $(V, \eta) $ is called a ``UJLA structure''.
\end{de}

\begin{re}
The UJLA structures  unify Jordan algebras, Lie algebras
and (non-unital) associative algebras; results for UJLA structures could
be ``decoded'' in properties of  Jordan algebras, Lie algebras
or (non-unital) associative algebras. This unification resembles the properties of quantum computers.
\end{re}

\begin{re}
The UJLA structures unify in a natural manner the above mentioned non-associative structures.
Thus, if $ (A, \ \theta )$, where $ \theta : A \otimes A \rightarrow A, \ \ 
\theta (a \otimes b) = ab $, 
 is a (non-unital) associative algebra, then we define  $ (A, \ \theta' )$, where $ \theta' (a \ot b) = \alpha ab \ + \ \beta ba $,
for $ \ \alpha, \ \beta \in k$.

If $\alpha = \frac{1}{2}$ and $ \beta = \frac{1}{2}$, then $ (A, \ \theta' )$ is a Jordan algebra.
%the opposite algebra of $ (V, \ \theta )$.

If $\alpha = 1$ and $ \beta = -1$, then $ (A, \ \theta' )$ is a Lie algebra. 

If $\alpha = 0$ and $ \beta = 1$, then $ (A, \ \theta' )$ is the opposite algebra of $ (A, \ \theta )$,
 and if $\alpha = 1$ and $ \beta = 0$, then $ (A, \ \theta' )$ is the algebra  $ (A, \ \theta )$.

If we put no restrictions on $ \alpha$ and $\beta$, then  $ (V, \ \theta )$ is a UJLA structure.

\end{re}

\begin{te}
 For $V$ a $k$-space, $ f: V \rightarrow k$ a k-map,  $ \ \alpha, \ \beta \in k$, and $e \in V$ such that $f(e)=1$,
the following structures can be associated.

(i) $ (V, M, e)$, a unital associative algebra, where
$M(v \ot w)= f(v)w+vf(w)-f(v)f(w)e$;  

(ii) $ (V, \ [,])$, a Lie algebra, where
$[v, \ w]= f(v)w-vf(w)$;

(iii) $ (V, \ \mu)$, a Jordan algebra, where
$\mu (v \ot w)= f(v)w+vf(w)$;

(iv) $ (V, M_{\alpha, \beta})$, a UJLA structure, where
$ M_{\alpha, \beta}
(v \ot w)= \alpha f(v)w+ \beta vf(w)$.  

\end{te}

\begin{Proof} (i)
 The proof is direct. We denote by `` $\cdot$ '' the operation $M$, in order to simplify our presentation.
We observe that `` $\cdot$ '' is commutative.
We first prove that $e$ is the unity of our algebra:
$ x \cdot e= f(x)e+ f(e)x- f(e)f(x)e= x = e \cdot x.$ 

Next, we prove the associativity of `` $\cdot$ '':
 $ (x \cdot y) \cdot z = f(x)f(y)z+f(x)f(z)y-f(x)f(y)f(z)e+ f(x)f(y)z+xf(y)f(z)-f(x)f(y)f(z)e-f(x)f(y)z$;

 $ x \cdot ( y \cdot z ) = f(x)f(y)z+ f(x)yf(z) - f(x)f(y)f(z)e+ xf(y)f(z) - f(x)f(y)f(z)e$.

It follows that $ (x \cdot y) \cdot z =x \cdot ( y \cdot z )$.

\bigskip

(ii) In this case

$ (x \cdot y) \cdot z = f(x)f(z)y - f(y)f(z)x$;

$ (y \cdot z) \cdot x = f(y)f(x)z - f(z)f(x)y$;

$ (z \cdot x) \cdot y = f(z)f(y)x - f(x)f(y)z$.

The Jacobi identity is verified.

\bigskip

We leave the cases (iii) and (iv) to be proved by the reader. 
\end{Proof}

\begin{re}
The above theorem  produces new examples of non-associative structures, and it  finds common information encapsulated in
these non-associative structures.
\end{re}

\begin{te}
Let $(V, \eta) $ be a UJLA structure, and $ \ \alpha, \ \beta \in k$. Then,
 $(V, \eta'), \ \ \eta'(a\ot b)= \alpha ab+ \beta ba $ is a
 UJLA structure.

\end{te}

\bigskip

\begin{re}
 The classification of UJLA structures is an open problem, and it
is more difficult than the problem of
classifying associative algebras (which is an open problem for higher dimensions).
\end{re}

\begin{te}
Let $V$ be a vector space over the field $k$, and $p,q \in k$.
For $f,g : V \rightarrow V $, we define $ M (f \ot g) = f * g = f *_{p,q} g = p f \circ g - q g \circ f: V \rightarrow V$.
Then:

(i) $ ( End_k (V), \ *_{p,q}) $ is a UJLA structure $ \  \forall p,q \in k$.

(ii) For $ \phi : End_k (V) \rightarrow End_k (V \ot V)$ a morphism of UJLA structures (i.e., $ \ \phi (f*g)= \phi(f) * \phi(g) \ $),
$ \ W= \{ f: V \rightarrow V \vert f \circ M = M \circ \phi (f) \}$ is a sub-UJLA structure of the structure defined at (i). In other words,
$ \ f*g \in W, \ \forall f, g \in W $.

\end{te}

\bigskip

\section{Conclusions and further implications}

Surveying topics
from abstract algebra to
computational methods, and 
from computer science to number theory,
the current
paper relates these subjects by unifying theories 
and the celebrated Yang-Baxter equation. We present new results
(theorems 3.2, 3.3, 3.5, 4.1, 6.4, 6.6 and 6.8), and we propose several open
problems and new interpretations.

Unifying the main non-associative structures,
the UJLA structures are structures
which resemble the properties of quantum computers.
Quantum computers could help in solving hard problems in
number theory and optimization theory, because they have
large computational power. For example,
such a computer could use the Shor's algorithm
to break algorithms encryption.

From some solutions of the Yang-Baxter equation, one could construct abstract
universal gates
from quantum computing. We explained how this equation is related to computer programming,
we studied problems about transcendental numbers, 
and we used computational methods in order to solve problems
related
to these topics.

Related to the equation 
(\ref{ybco}) there is a long standing open problem.
 The following system of equations, obtained  
in \cite{dpfn} and extended in \cite{hin}, is not completely classified: 

\begin{eqnarray} &&
(\beta(v,w)-\gamma(v,w))(\alpha(u,v)\beta(u,w) -
\alpha(u,w)\beta(u,v))\nonumber \\ &&\quad \quad \quad +
(\alpha(u,v)-\gamma(u,v))(\alpha(v,w)\beta(u,w) - \alpha(u,w)\beta(v,w))
= 0 \label{e1} \\ \nonumber \\ &&
\beta(v,w)(\beta(u,v)-\gamma(u,v))(\alpha(u,w)-\gamma(u,w)) \nonumber \\
&&\quad \quad \quad +
(\alpha(v,w)-\gamma(v,w))(\beta(u,w)\gamma(u,v)-\beta(u,v)\gamma(u,w)) =
0 \label{e2} \\ \nonumber \\ && \alpha(u,v)
\beta(v,w)(\alpha(u,w)-\gamma(u,w)) + \alpha(v,w)\gamma(u,w)
(\gamma(u,v) - \alpha(u,v)) \nonumber \\ &&\quad \quad \quad +
\gamma(v,w) (\alpha(u,v)\gamma(u,w)-\alpha(u,w)\gamma(u,v)) = 0
\label{e3} \\ \nonumber \\ && \alpha(u,v)
\beta(v,w)(\beta(u,w)-\gamma(u,w)) + \beta(v,w)\gamma(u,w) (\gamma(u,v)
- \beta(u,v)) \nonumber \\ &&\quad \quad \quad + \gamma(v,w)
(\beta(u,v)\gamma(u,w)-\beta(u,w)\gamma(u,v)) = 0 \label{e4} \\
\nonumber \\ && \alpha(u,v)(\alpha(v,w)-\gamma(v,w))(\beta(u,w) -
\gamma(u,w)) \nonumber \\ &&\quad \quad \quad + (\beta(u,v)-
\gamma(u,v))( \alpha(u,w) \gamma(v,w) - \alpha(v,w) \gamma(u,w)) = 0 \
\label{e5} \end{eqnarray}

From the transdisciplinary (see \cite{b, fn}) point of view,
and attempting to relate art and science, the equation (\ref{ecgraddoi})
could be called the ``cubism equation''. In the same manner, the inverse system 
could be related to Art Nouveau (for example, recall the architecture of Casa Mila by Gaudi).
%\ifCLASSOPTIONcaptionsoff
%  \newpage
%\fi

\bigskip

\section*{Acknowledgment}

The author would like to thank  the organizers of the
INASE Conference in Barcelona (April 7-9, 2015), for inviting us to present our results at that conference,
and to the
Simion Stoilow Institute of Mathematics
of the Romanian Academy for some support.

\end{document}